\documentclass[conference]{IEEEtran}

\usepackage[letterpaper, left=0.680in, right=0.680in, bottom=1.049in, top=0.71in]{geometry}

\usepackage{ifpdf}
\ifCLASSINFOpdf
\usepackage[pdftex]{graphicx}
\graphicspath{{./Figures/}}
\usepackage[caption=false, font=normalsize, labelfont=sf, textfont=sf]{subfig}
\DeclareGraphicsExtensions{.pdf,.jpeg,.png}
\else
\usepackage[dvips]{graphicx}
\DeclareGraphicsExtensions{.pdf}
\usepackage[caption=false, font=footnotesize]{subfig}
\fi

\usepackage{subfig}
\usepackage{caption}

\usepackage{cite}
\usepackage{amsmath,amssymb,amsfonts}
\usepackage{algorithmic}
\usepackage{graphicx}
\usepackage{textcomp}
\usepackage{xcolor}
\definecolor{royalblue}{HTML}{4169E1}
\definecolor{forestgreen}{HTML}{228B22} 
\definecolor{firebrick}{HTML}{B22222} 
\usepackage{svg}

\def\BibTeX{{\rm B\kern-.05em{\sc i\kern-.025em b}\kern-.08em
    T\kern-.1667em\lower.7ex\hbox{E}\kern-.125emX}}

\begin{document}

\title{Domain Adaptation-Enabled Realistic Map-Based Channel Estimation for MIMO-OFDM}

\author{\IEEEauthorblockN{
            Thien~Hieu~Hoang\IEEEauthorrefmark{1},
            Tri~Nhu~Do\IEEEauthorrefmark{2},
            and~Georges~Kaddoum\IEEEauthorrefmark{1}%
			}
		\IEEEauthorblockA{
		\IEEEauthorrefmark{1} Department of Electrical Engineering, \'{E}TS, Montr\'{e}al, Qu\'{e}bec, Canada\\
        \IEEEauthorrefmark{2}Department of Electrical Engineering, Polytechnique Montr\'{e}al, Montr\'{e}al, Qu\'{e}bec, Canada
        }
		\IEEEauthorblockA{ Emails: 
			thien-hieu.hoang.1@ens.etsmtl.ca,
			tri-nhu.do@polymtl.ca,
			Georges.Kaddoum@etsmtl.ca
			}
	}

\maketitle

\begin{abstract}
Accurate channel estimation is crucial for the improvement of signal processing performance in wireless communications. 
    However, traditional model-based methods frequently experience difficulties in dynamic environments.
Similarly, alternative machine-learning approaches typically lack generalization across different datasets due to variations in channel characteristics.
    To address this issue, in this study, we propose a novel domain adaptation approach to bridge the gap between the quasi-static channel model (QSCM) and the map-based channel model (MBCM). Specifically, we first proposed a channel estimation pipeline that takes into account realistic channel simulation to train our foundation model. Then, we proposed domain adaptation methods to address the estimation problem.
Using simulation-based training to reduce data requirements for effective application in practical wireless environments, we find that the proposed strategy enables robust model performance, even with limited true channel information.
\end{abstract}


\section{Introduction}
Accurate channel estimation, which has a direct impact on the performance of various signal processing algorithms, such as beamforming,
detection, and equalization, is a critical task in wireless communications. 
    With the growth in demand for higher data rates and more reliable communications, channel estimation has garnered significant attention in both academic research and industrial applications.
Traditional approaches such as least squares and linear minimum mean square error have been widely adopted; however, considering the complexity of modern communication systems, these methods frequently fall short in dynamic or high-dimensional environments \cite{Wei2021}.

In recent years, machine learning, which offers data-driven solutions
that can adapt to complex environments, has emerged as a promising tool for channel estimation. 
    By leveraging large amounts of data, machine-learning models can learn to more effectively approximate the behavior of wireless channels than conventional methods \cite{Elbir2022}. 
Nonetheless, machine-learning-based models are typically trained on a specific dataset, and their generalization ability may be compromised when applied to different environments \cite{Farahani2021}. 
    This discrepancy arises due to distribution shifts in channel characteristics between datasets, necessitating the use of domain adaptation techniques. %

Domain adaptation plays a crucial role in scenarios where a model trained on one dataset does not perform adequately on another dataset due to differences in the underlying data distributions. 
    In practical wireless communications, obtaining real-world channel data for training is frequently limited by high costs, privacy concerns, or unavailability of the true channel \cite{Akrout2023}.
This creates a significant challenge - namely, how to train robust models in the face of limited data from the real world.

    {This study proposes a pipeline that leverages simulated data for training before deploying the model in real-world scenarios, reducing dependence on true channel data. 
However, models trained solely on simulations often struggle due to distribution shifts. 
    To address this, we apply domain adaptation to bridge the gap between the simulated data and a semi-practical map-based scenario with real-world characteristics. 
This approach enhances generalization, enabling effective performance in practical settings with limited real channel data.
    The key contributions of this research can be summarized as:
    i) A domain adaptation framework for channel estimation is developed to bridge the Quasi-Static Channel Model (QSCM) and the Map-Based Channel Model (MBCM).
    ii) CNN and GAN models are integrated with transfer learning to enhance generalization across different environments.
    iii) A realistic, semi-practical dataset is constructed using OpenStreetMap and MATLAB RayTracing, incorporating real-world propagation characteristics to improve model training.
}

\begin{figure*}[t]
	\centering
    \subfloat[\textnormal]{%
        \includegraphics[width=.24\linewidth]{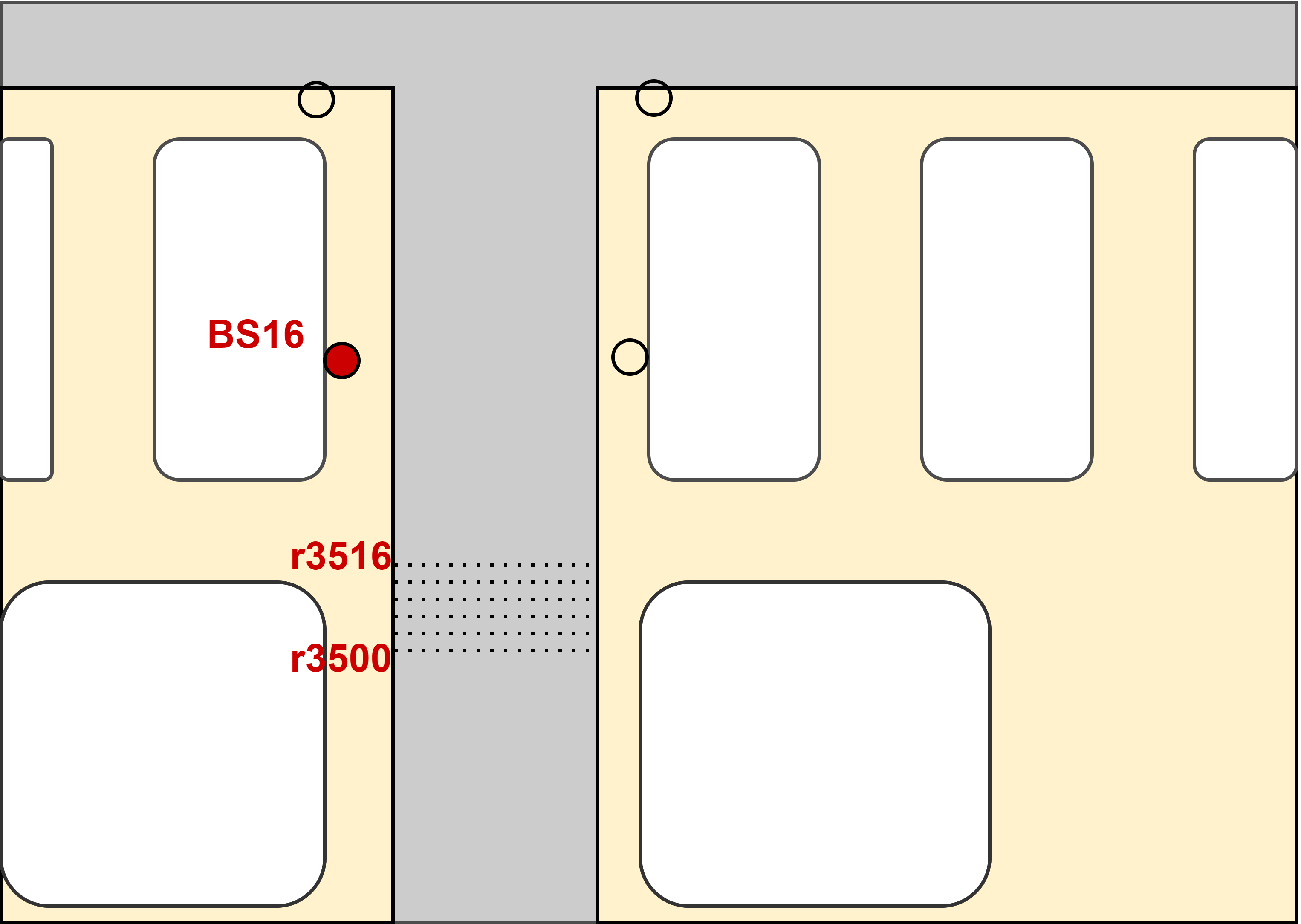}
        \label{fig:1_DeepMIMOmap}} \hfill
	\subfloat[]{%
		\includegraphics[width=.24\linewidth]{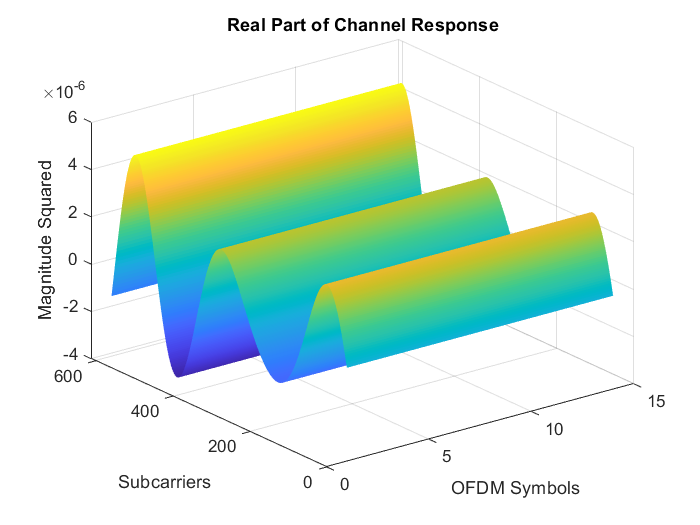}
		\label{fig:1_DeepMIMOChan}} \hfill
	\subfloat[]{%
		\includegraphics[width=.24\linewidth]{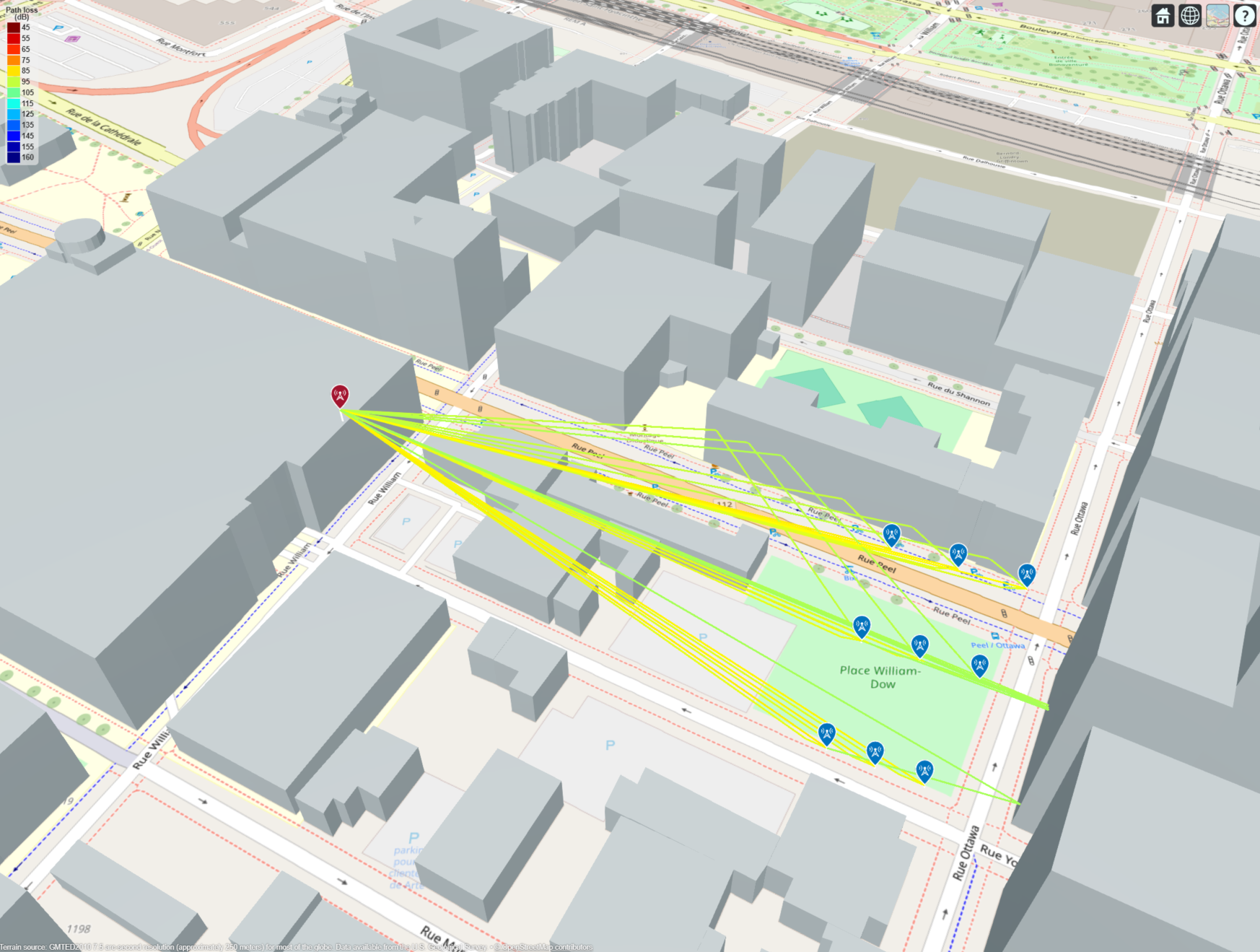}
		\label{fig:1_mapBase}} 
    \subfloat[]{%
		\includegraphics[width=.24\linewidth]{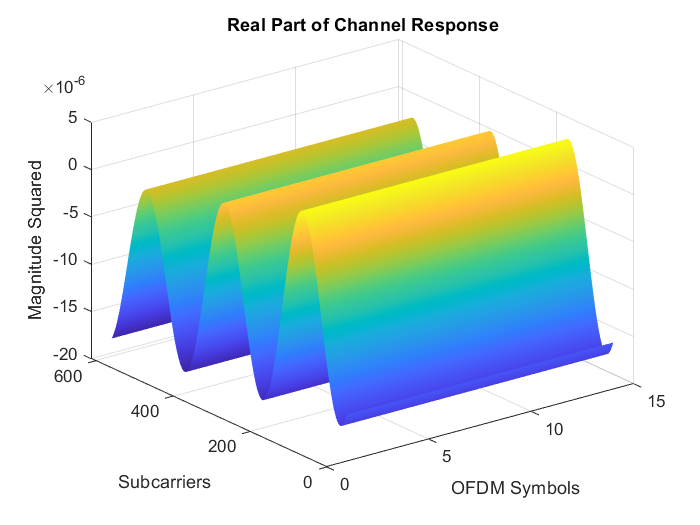}
		\label{fig:1_ETSChan}} 
	\caption{(a) Source domain: QSCM with DeepMIMO scenario O1; (b) Generated channel from QSCM; (c) Target Domain: MBCM with \'{E}TS neighborhood map-based scenario; (d) Generated channel from MBCM}
	\label{fig:scenarios} 
\end{figure*}

\section{System Model and Signal Modelling}

We consider a downlink point-to-point system consisting of one base station (BS) and one user equipment (UE).  
{ The antenna locations are represented by the position vector $\vec{p}_A, A\in \{\tt BS, UE\}$. 
    The BS's antenna position $\vec{p}_{\tt BS}$  remains fixed, while the UE's position $\vec{p}_{\tt UE}(i)$ varies in each iteration.}

\subsection{Quasi-Static Channel Model (QSCM) for Foundation Domain Analysis and Foundation Model training} \label{sec:quasiModel}
{In realistic 5G Orthogonal Frequency Division Multiplexing (OFDM) systems, the channel is influenced by factors such as multipath propagation, Doppler shifts, angle of arrival (DoA), angle of departure (DoD), non-line-of-sight (NLOS) conditions, and environmental variations, all of which impact channel estimation accuracy and require robust domain adaptation techniques. 
    These dynamic channel characteristics necessitate adaptive models capable of handling diverse real-world scenarios, including mobility and hardware impairments.
Therefore, our simulation environment must accurately capture these realistic channel characteristics. 
    For this reason, we selected DeepMIMO \cite{Alkhateeb2019} as our simulation dataset.}
We consider a block fading channel \cite{Jiang2022}, where the channel remains time-invariant within each slot, but varies from one slot to the next. 

Since the time-varying channel is primarily due to the Doppler effect \cite{Jiang2022},
  we assume a quasi-static time-invariant channel model by setting $f_{D_\ell}=0$ within one slot, where
     $f_{D_ \ell}$ is the $\ell$-th path Doppler frequency shift.
In the RG, one resource element --- the channel for the OFDM symbol $m $ $(1 \leq m \leq M)$ at subcarrier $k$ --- is represented as shown in Eq. \eqref{eq:H_km}, assuming no Doppler effect \cite{Alkhateeb2019}:
\begin{align}
    {H} ({k,m} ) & = \sum_{\ell=0}^ {L-1} h_{\ell, m}  e ^{j\left(\phi_{\ell, m} - \frac{2 \pi k}{K} B \tau_{\ell, m} \right)} \nonumber \\ 
    \times & \mathbf{a}_{\tt BS} \left( \phi_{\tt BS} ^{\mathrm{az}}, \phi_{\tt BS} ^{ \mathrm{el}}\right) 
    \mathbf{a} ^H_{\tt UE} \left(\phi_{\tt UE} ^{\mathrm{az}}, \phi_{\tt UE} ^{\mathrm{el}}\right), \ 
    0 \leq k \leq K-1,
    \label{eq:H_km}
\end{align}
where $L$ is the number of propagation paths,
$h_{\ell, m}$ is the complex impulse response of the $\ell$-th path,
    $\tau_{\ell, m}$ is the $\ell$-th path delay normalized by sampling time, and
  $\phi_{\ell, m}$ is the phase of path $\ell$.
Given two symbols, $m_1$ and $m_2$, in the same slot, we can assume that $\phi_{\ell, m_1} = \phi_{\ell, m_2}$, $\tau_{\ell, m_1} = \tau_{\ell, m_2}$.
    This, in turn, implies that     
    $H(k,m_1) = H(k,m_2), \forall m_1, m_2 \in \{1,...,14\}$.
Here, $\mathbf{a}_{\tt BS} \left(\phi_{\tt BS} ^{\mathrm{az}}, \phi_{\tt BS} ^{\mathrm{el}}\right)$ and 
    $\mathbf{a}_{\tt UE} \left(\phi_{\tt UE} ^{\mathrm{az}}, \phi_{\tt UE} ^{\mathrm{el}}\right)$ are the phase responses of BS and UE, respectively.
The phase response $\mathbf{a}_{\tt BS} \left(\phi_{\tt BS} ^{\mathrm{az}}, \phi_{\tt BS} ^{\mathrm{el}}\right)$ of the BS is given by 
    $\mathbf{a}_{\tt BS} \left( \phi_{\tt BS} ^{ \mathrm{az} },  \phi_{\tt BS} ^{ \mathrm{el} }\right) = [1, e^{j kd \sin (\phi_{\tt BS} ^{\mathrm{az}}) \cos( \phi_{\tt BS} ^{\mathrm{el}})}, ...,
    e^{j kd (M _{\tt BS}-1) \sin (\phi_{\tt BS} ^{\mathrm{az}}) \cos( \phi_{\tt BS} ^{\mathrm{el}})}]^{\mathrm{T}}$, 
where $M _{\tt BS}$ is the antenna number of BS, 
  $d$ represents the spacing between antenna elements when the BS antennas are arranged in an array,
$\phi_{\tt BS} ^{\mathrm{az}}$, $\phi_{\tt BS} ^{\mathrm{el}}$ are the azimuth and elevation angles of arrival at the BS, respectively.
    In our point-to-point system, since $M _{\tt BS}=1$, we obtain $\mathbf{a}_{\tt BS} \left( \phi_{\tt BS} ^{ \mathrm{az} },  \phi_{\tt BS} ^{ \mathrm{el} }\right) = 1$. 
Similarly, we have $\mathbf{a}_{\tt UE} \left( \phi_{\tt UE} ^{ \mathrm{az} },  \phi_{\tt UE} ^{ \mathrm{el} }\right) = 1$. 

The positions of the BS and UEs are shown in Fig. \ref{fig:1_DeepMIMOmap}, while a sample of generated channels is illustrated in Fig. \ref{fig:1_DeepMIMOChan}.
    We refer to this dataset as the quasi-static model (QSCM) data that serve as the source domain in our channel estimation transfer learning task.

\subsection{Map-Based Channel Model (MBCM) for Realistic AI Model Deployment} \label{sec:mapMode}

We use MATLAB's RayTracing object to generate a semi-practical channel dataset for MBCM.
 The channel in RayTracing is generated using the Cluster Delay Line (CDL) model and represented in the time-domain sampling expression.
    The channel for OFDM symbol $m$ at sampling index $n$, denoted by $h(n,m)$, 
     is given by Eq. \eqref{eq:h_n_m1}.
\begin{align}
    h(n,m) = \sum_{\ell =0}^{L-1} & h_{\ell, m} e^{j 2 \pi f_{D_ \ell} \frac{n}{N}} 
     e^{j \phi_{\ell, m}}
       \delta\left( n -\tau_{\ell, m} \right) \nonumber \\ 
    & \times \mathbf{a}_{\tt BS} \left( \phi_{\tt BS} ^{ \mathrm{az}}, \phi_{\tt BS} ^{ \mathrm{el} } \right) \mathbf{a} ^{H}_{\tt UE} \left( \phi_{\tt UE} ^ { \mathrm{az}}, \phi_{\tt UE} ^{ \mathrm{el}} \right), \label{eq:h_n_m1}
\end{align}
{where $N=K$ is the number of samples representing one OFDM symbol and  
     $n$ ($0 \leq n \leq N-1$) denotes the sample index.
Note that $\{ h(n, m) \}_{n=0} ^ {N-1}$ here is the  Inverse Discrete Fourier Transform (IDFT) of $\{ H(k, m) \}_{k=0} ^ {K-1}$ presented in \eqref{eq:H_km} regarding $f_{D_{\ell}} =0$.}

On obtaining the channels in time-domain sampling representation from RayTracing, we convert them back to the channel grid, i.e., subcarrier-OFDM symbol representation. 
    These channels in the MBCM dataset serve as our target domain dataset in our channel estimation transfer learning problem.
Fig. \ref{fig:1_mapBase} provides an example of the positions of the BS and UEs, while Fig. \ref{fig:1_ETSChan} shows an example generated channel in our MBCM scenario.

\subsection{Practical DM-RS-based Downlink Channel Estimation}

\begin{figure*}[htb]
    \centering
    \includegraphics[width=.9\linewidth]
    {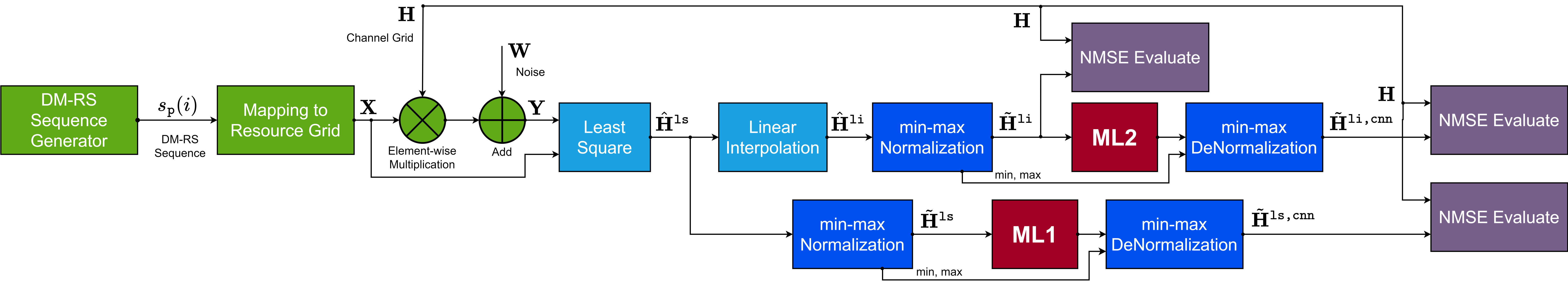}
    \caption{System architecture of ML-based channel estimation.}
    \label{fig:4_diagram2}
\end{figure*}

\subsubsection{DM-RS Pilot}

The DM-RS (Demodulation Reference Signal) sequence $s_{\tt p}$  in 5G-NR is generated 
  as described in Section 7.4 in \cite{TS.38.211}.
The sequence is then mapped to the RG to get the transmit RG $\mathbf{X}$.
    The mapping positions are defined in Section 7.4. 
    Our setting is ``configuration type 1" and ``mapping type A".
The $i$-th pilot in the pilot sequence is the $i$-th element of the pilot elements,
    $X(k_{\tt p}^i,m_{\tt p}^i)=s_{\tt p}(i)$;
the other elements of $\mathbf{X}$ are zeros.
    The $m$-th column of $\mathbf{X}$, $\{X(k,m) \}_{k=0} ^{K-1}$, represents the frequency-domain presentation of the $m$-th OFDM symbol. 

\subsubsection{OFDM Waveform}

The transmit RG is then passed through an IDFT, and each OFDM symbol $\{ X(k,m) \}_{k=0} ^{K-1}$ is transformed into the corresponding OFDM time-domain waveform $\{ x(n,m) \}_{n=0} ^{N-1}$ as (see Eq. \eqref{eq:x_n_m}; \cite{Han1998}).  
\begin{align}
    x(n,m) \! = \! \operatorname{IDFT}\{X(k,m)\} \! = \! \frac{1}{N} \sum_{\substack{k=0}}^{N-1} & X(k,m) e^{j 2 \pi k n / N}. 
    \label{eq:x_n_m}
\end{align}

The cyclic prefix (CP) is then inserted into the waveform to prevent inter-symbol interference (ISI) during transmission and reception. 
    We get the OFDM symbol with CP $\{ x_{\tt cp}(n, m) \}_{n= -N_{\tt cp}} ^{N -1}$ of OFDM symbol $m$ as shown in Eq. \eqref{eq:x_cp_nm}.
\begin{align}
    x_{\tt cp}(n,m)= 
    \begin{cases}
        x(N+n, m), & -N_{\tt cp} \leq n \leq -1 \\ 
        x(n, m),   & 0 \leq n \leq N-1, \label{eq:x_cp_nm}
    \end{cases}
\end{align}
where $N_{\tt cp}$ is the length of the CP. 

    The transmitted signal is then sent to a multi-path fading channel and the received $m$-th OFDM waveform $\{ {y}_{\tt cp}(n, m) \}_{n= -N_{\tt cp}} ^ {N -1}$  can be represented by
    ${y}_{\tt cp}(n,m) = x_{\tt cp}(n,m) \otimes h(n, m) + w(n,m)$, 
where $\otimes$ denotes the convolution operator along $n$ dimension, $h(n, m)$ is the channel impulse response denoted in Eq. \eqref{eq:h_n_m1}, and $w(n,m)$ represents the additive white Gaussian noise.

    On removing CP, we get received samples $\{ {y}(n,m) \}_{n=0} ^{N-1}$ of OFDM symbol $m$, then $\{ {y}(n,m) \}_{n=0} ^{N-1}$ are sent to a Discrete Fourier Transform (DFT) block to demultiplex the multicarrier signals,
    ${Y}(k,m)=\operatorname{DFT}\{{y}(n,m)\}$.
{With a sufficiently long CP guard, ISI impact is significantly reduced \cite{Han1998},
    and the demultiplexed symbols $\{ {Y}(k,m) \}_{k=0} ^{K-1}$ are given by Eq. \eqref{eq:Y=XH+W}.}
\begin{align}
    {Y}(k,m) = X(k,m) H(k,m) + W(k,m), 
    \label{eq:Y=XH+W}
\end{align}
where $W(k,m)$ is the IDFT of $w(n,m)$.

\subsection{Domain Adaptation Problem Formulation}

The key challenge in domain adaptation is handling the distribution shift between the source and target domains, which leads to poor generalization if straightly applied.
Specifically, we define the source domain dataset as $\mathcal{D}_{\tt T} = \{\mathbf{H}_{\tt S}^i \}_{i=1} ^{N_{\tt S}}$ and target domain dataset as $\mathcal{D}_{\tt S} = \{ \mathbf{H}_{\tt T}^i \}_{i=1} ^{N_{\tt T}}$, where 
    $N_{\tt S}$, $N_{\tt T}$ are the number of data samples in the source and target domains, respectively. 
  In this context, $\mathbf{{H}}_{\tt S}^i $, $\mathbf{{H}}_{\tt T}^i $ represent the true channels, serving as the labels for our machine-learning models.  
    Our target is to estimate the true channel $\mathbf{{H}}$ from the available information $\mathbf{\tilde{H}}$ using machine-learning models. 
  Here, $\mathbf{\tilde{H}}$ refers to either normalized least squares (LS)-based channel estimates or normalized least squares with linear interpolation (LS-LI)-based channel estimates (see Section \ref{sec:source} for further discussion).
Since the available data in our semi-practical scenario (the target domain) are limited, we aim to pre-train our model using data from the QSCM simulation (the source domain).
    However, the joint distributions of the input data $\mathbf{\tilde{H}}$ and the true channel $\mathbf{H}$ differ between these two domains, i.e., ${P}_{\tt S} (\mathbf{\tilde{H}}, \mathbf{H}) \neq P_{\tt T} (\mathbf{\tilde{H}}, \mathbf{H})$,
        which leads to a poor performance of our pre-trained model in the target domain. 
    Therefore, the challenge of our domain adaptation problem is to manage the distribution shift between these two domains.

Our problem is a semi-supervised domain adaptation problem, where we have labeled data in the source domain but only a limited amount of labeled data in the target domain; our goal is to leverage both labeled and unlabeled target data to effectively transfer knowledge from the source domain to the target domain. 
    The general optimization problem can be written as 
$
\min_{\theta} \left( \mathcal{L}_{\tt S}(\theta_{\tt S}) + \lambda \mathcal{L}_{\tt T}(\theta_{\tt T}) \right),
$
where 
    $\mathcal{L}_{\tt S}(\theta_{\tt S}) = \mathbb{E}_{( \mathbf{ \tilde{H}}_{\tt T}, \mathbf{H}_{\tt T}) \sim P_{\tt T}} \left[ \ell(f(\mathbf{ \tilde{H}}_{\tt S}; \theta_{\tt S}), \mathbf{H}_{\tt S}) \right]$  
and 
    $\mathcal{L}_{\tt T}(\theta_{\tt T}) = \mathbb{E}_{( \mathbf{ \tilde{H}}_{\tt T}, \mathbf{H}_{\tt T}) \sim P_{\tt T}} \left[ \ell(f( \mathbf{ \tilde{H}}_{\tt T}; \theta_{\tt T}), \mathbf{H}_{\tt T}) \right]$
  are the losses on the source and target domains, respectively,
$\lambda$ is the regularization parameter controlling the balance between source and target domain losses,
    $\ell ( \cdot )$ is the loss function, 
      $f( \mathbf{ \tilde{H}}_{\tt S}; \theta_{\tt S})$ and $f( \mathbf{ \tilde{H}}_{\tt T}; \theta_{\tt T})$ are the model’s predictions for the source and target domains, using parameters $\theta_{\tt S}$ and $\theta_{\tt T}$, respectively.
{In this work, the domain adaptation optimization problem is applied to our proposed hybrid framework for channel estimation on QSCM and MBCM. 
    To address this challenge, transfer learning is employed as a key strategy, and the details of this approach will be discussed in detail in Sections \ref{sec:source} and \ref{sec:target}.
}

\section{The Proposed Hybrid Estimation Framework for Source Model Training} \label{sec:source}


The framework integrates conventional estimation methods with machine-learning models to improve channel estimates by correcting the inaccuracies in the initial Least Squares (LS) and Least Squares-Linear Interpolation (LS-LI) estimates.
    The training framework consists of two main modules: 
i) Main Channel Estimation Module. This initial module employs traditional algorithms, including LS estimator and LS-LI estimator, to generate preliminary channel matrix estimates denoted by $\hat{\mathbf{H}} ^{\tt ls}$ and $\hat{\mathbf{H}} ^{\tt li}$, respectively. 
ii) Machine-Learning Module. The second module focuses on enhancing estimation accuracy of  $\hat{\mathbf{H}} ^{\tt ls}$ and $\hat{\mathbf{H}} ^{\tt li}$  by treating the problem as one of imperfect Channel State Information (CSI). 
    In this module, Convolutional Neural Networks (CNNs) and Generative Adversarial Networks (GANs) are used to improve quality of the estimated channel matrices. 
The machine-learning process aims to rectify imperfections inherent in the LS-based and LS-LI-based estimates, ultimately yielding more precise channel estimations.

\subsection{Least Square Estimation}
    
The Least-Squared (LS) estimated channel $\mathbf{\hat{H}}^{\tt ls}$ response at the pilot positions $(k_p, m_p)$ can be estimated by
    $\hat{H}^{\tt ls}(k_p, m_p) = {{Y}(k_p, m_p)} / {X(k_p, m_p)}$.
The remaining positions in the channel grid are set to zero.
    Consequently, we obtain the LS channel estimate denoted by $\mathbf{\hat{H}}^{\tt ls}$.
This $\mathbf{\hat{H}}^{\tt ls}$ plays a crucial role in this study.
    On obtaining $\mathbf{\hat{H}}^{\tt ls}$,  we apply linear interpolation to reconstruct the complete channel grid.
The resulting channel after linear interpolation is denoted as $\mathbf{\hat{H}}^{\tt li}$.

\subsection{Convolutional Neural Network Approach}

Before feeding $\mathbf{\hat{H}}^{\tt ls}$ and $\mathbf{\hat{H}}^{\tt li}$ to the CNNs, we use min-max scaler $S$ to normalize them to the range $[-1, 1]$,
    $\mathbf{\tilde{H}}^{\tt ls} = S ( \mathbf{\hat{H}}^{\tt ls} ), 
    \mathbf{\tilde{H}}^{\tt li} = S ( \mathbf{\hat{H}}^{\tt li} )$.
We use two CNN models to improve the estimation performance of the least squares and linear interpolation approaches.
    These networks aim to correct the imperfection in the LS-based, $\mathbf{\hat{H}}^{\tt ls}$, and LS-LI-base estimates, $\mathbf{\hat{H}}^{\tt li}$, thereby providing a more accurate estimation of the channel.
We denote CNN1 as the model that uses the data-label pairs $( \mathbf{\tilde{H}}^{\tt ls},  \mathbf{{H}} )$, 
    and CNN2 as the model that uses $( \mathbf{\tilde{H}}^{\tt li},  \mathbf{{H}} )$ as its data-label pairs.
The diagram of our system is shown in Fig. \ref{fig:4_diagram2}, where ML1 and ML2 correspond to CNN1 and CNN2, respectively.
{The first layer of the network is the batch normalization $g_0 ^{\tt bnorm}$.
In $i$-th layer block of CNN, we design convolution layer $g_i ^{\tt conv}$, `Tanh' activation $g_i ^{\tt act}$.
    Convolution layer $g_0 ^{\tt conv}$ has kernel size $(9,9)$, stride $1$, padding $4$; 
convolution layers $g_1 ^{\tt conv},..., g_6 ^{\tt conv}$ are with kernel size $(5,5)$, stride $1$, padding $2$.
    The output of the CNN is 
    $( g_6^{\tt conv} \circ g_5^{\tt act} \circ g_5^{\tt conv} \circ \cdots  \circ g_0^{\tt act} \circ g_0^{\tt conv} \circ g_0^{\tt bnorm}  ) (\mathbf{\Tilde{H}}) $, where $\mathbf{\Tilde{H}}$ can be $\mathbf{\Tilde{H}} ^{\tt ls}$ or $\mathbf{\Tilde{H}} ^{\tt li}$.
  The numbers of filters from layer 0 to layer 6 are in $[64,64,64,32,16,8,1]$, respectively.}
Since the models are trained on the QSCM dataset,
    we refer to these models as the QSCM models.

\subsection{Generative Adversarial Network Approach}
For our GAN approach, we employ Pix2Pix \cite{Isola2017}, which is specifically designed for image-to-image translation tasks. 
    The GAN architecture consists of the following two key models: a generator and a discriminator.
In the Pix2Pix generator, we use a UNet architecture consisting of 7 encoder layers $f_{i, \tt enc} ^{\tt G}$, and 7 decoder layers $f_{i, \tt dec} ^{\tt G}$, $(i \in \{1,\dots,7 \})$.
  Skip connections are employed to concatenate or add feature maps from the downsampling path to the corresponding layers in the upsampling path.
A bottle-neck layer $f^{\tt G} _{\tt bottle}$ connects the encoder and decoder paths. 
    In the Discriminator, we use $5$ convolutional layers, $f^{\tt D}_i \ (i \in \{1,...,5\})$.

Similarly to CNNs, we denote GAN1 as the model using the data-label pairs $( \mathbf{\tilde{H}}^{\tt ls},  \mathbf{{H}} )$, 
    and GAN2 as the model that uses $( \mathbf{\tilde{H}}^{\tt li},  \mathbf{{H}} )$ as its data-label pairs.
In this context, the ML1 and ML2 blocks in the diagram in Fig. \ref{fig:4_diagram2} represent  GAN1 and GAN2, respectively.

\subsection{Evaluation Approach}
After de-normalizing the results from ML networks with $\min$, $\max$ of $\mathbf{\hat{H}}^{\tt ls}$ and $\mathbf{\hat{H}}^{\tt li}$, we get estimated channels $\mathbf{\hat{H}}^{\tt ls, cnn}$ and $\mathbf{\hat{H}}^{\tt li, cnn}$, respectively.
    To evaluate the estimated channels, we use normalized mean squared error (NMSE) to compare with the actual channel $\mathbf{H}$,
    $NMSE = \frac{ || \mathbf{\hat{H}} - \mathbf{H} || _2 ^2 }{ ||\mathbf{H} || _2 ^2}$,
where $\mathbf{\hat{H}}$ is the estimated channel from the CNN approach, $\mathbf{\hat{H}} ^{\tt ls, cnn}$, $\mathbf{\hat{H}} ^{\tt li, cnn}$,
  or from the GAN approach, $\mathbf{\hat{H}} ^{\tt ls, gan}$, $\mathbf{\hat{H}} ^{\tt li, gan}$.
    NMSE is also used to evaluate the result of the LS-LI approach, $\mathbf{\hat{H}} ^{\tt li}$.

\section{Domain Adaptation for Estimation of MBCM} \label{sec:target}

In this study, we tackle the formulated domain adaptation problem using fine-tuned transfer learning. 
    Specifically, to improve the model's ability to generalize to the target domain, we take the pre-trained models from the source domain (the QSCM dataset) and fine-tune them on the target domain (the MBCM dataset).
 The mismatch between the source and target domains is characterized by a shift in channel distribution from QSCM to MBCM (from quasi-stochastic to quasi-deterministic) as shown in Fig.~\ref{fig:3_dist_shift}.

\subsection{Data in Target Domain}

\begin{figure}
    \centering
    \subfloat[]{%
        \includegraphics[width=.425 \linewidth]{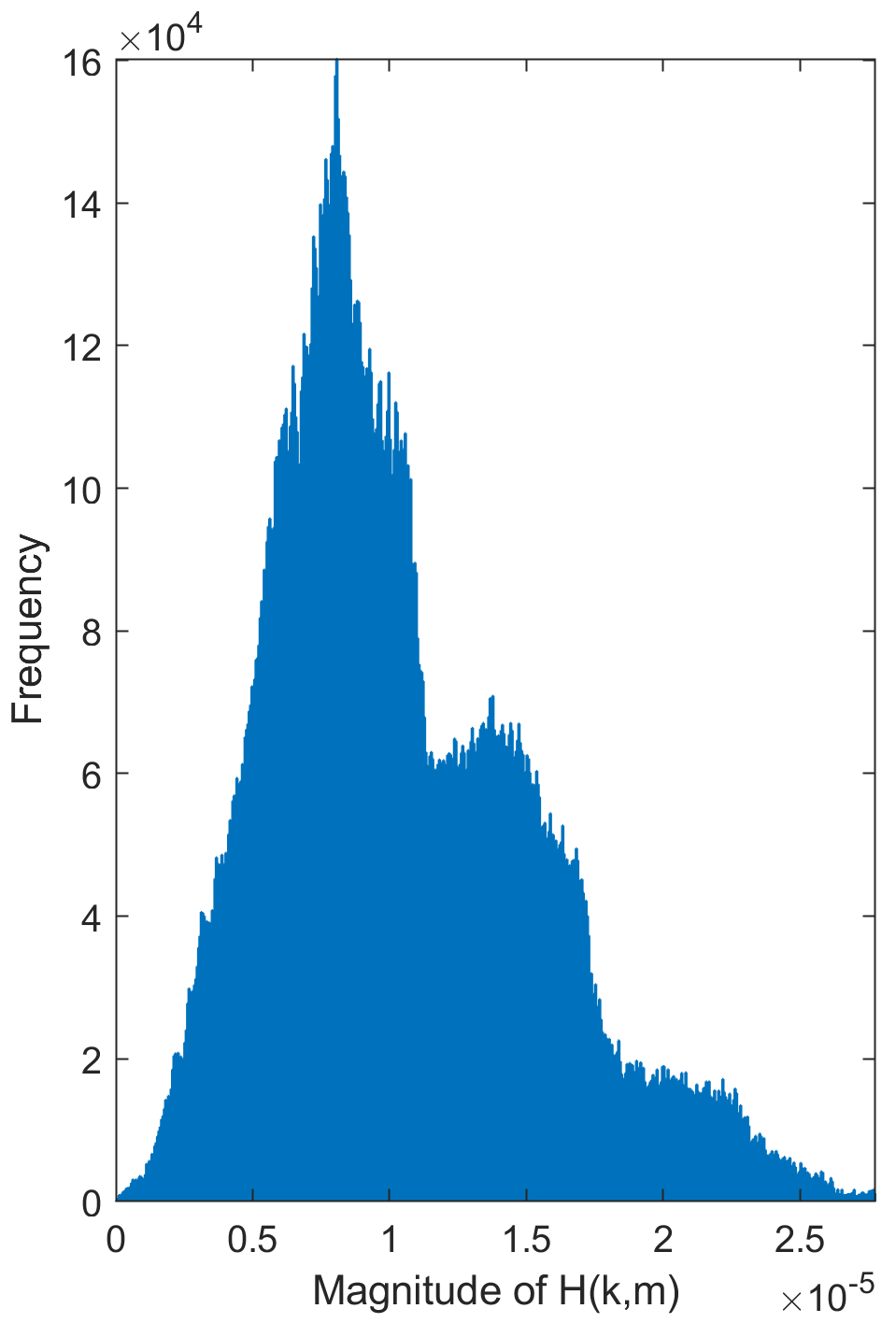}
        \label{fig:3_dist_DeepMIMO}} \hfill
	\subfloat[]{%
		\includegraphics[width=.46\linewidth]{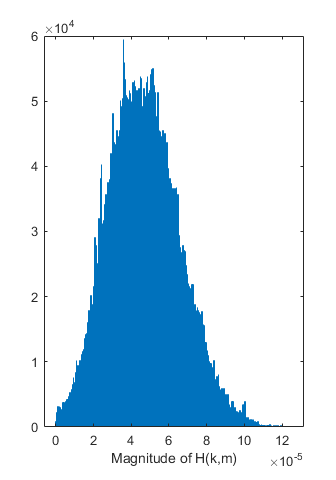}
		\label{fig:3_dist_MapBase}} \hfill
    \caption{Distribution shift of magnitudes of $H(k,m)$ in (a) source domain: QSCM dataset, and (b) target domain: MBCM dataset.}
    \label{fig:3_dist_shift}
\end{figure}

Differences in the locations of BSs and UEs, as well as the surrounding building environments, result in distinct parameter distributions between the two datasets. 
    These variations include the number of paths $L$, path gains $h_{\ell, m}$, path phases $\phi_{\ell, m}$, path delays $\tau_{\ell, m}$, and angles of arrival and departure.
Consequently, these variations influence the magnitudes of $H(k,m)$, which can be observed in its histogram at 0 dB SNR (see Fig. \ref{fig:3_dist_shift}).
    The magnitudes of our data in the two domains have distinct ranges --- one is $[0, 3\times 10^{-5}]$ and the other is in $[0, 12 \times 10^ {-5}]$ --- and different distributions.
We use the Wasserstein distance to measure the difference between the two distributions instead of Kullback-Leibler and Jensen-Shannon divergences, as our datasets have different ranges and large non-overlapping regions. 
    Furthermore, it captures the geometric structure of distributions and is more robust to small variations. 
Mathematically, the Wasserstein-1 distance between our two datasets 
can be given as shown in Eq. \eqref{eq:W_1_d}.
\begin{align}
    W_1\left(\mathcal{D}_{\tt {S }}, \mathcal{D}_{\tt {T }}\right)=\min _\gamma \sum_{i=1}^{N_{\tt S}} \sum_{j=1}^{N_ {\tt T}} \gamma_{i j}\left\|    \mathbf{H}_{{\tt S}} ^i -\mathbf{H}_{{\tt T}} ^j\right\|  _F ,
    \label{eq:W_1_d}
\end{align}
subject to
    $\sum_{j=1}^{N_{\tt T}} \gamma_{i j}=\frac{1}{N_{\tt S}} \quad$ for all $i \quad\left(i=1, \ldots, N_{\tt S}\right)$,
$\sum_{i=1}^{N_{\tt S}} \gamma_{i j}=\frac{1}{N_{\tt T}} \quad$ for all $j \quad\left(j=1, \ldots, N_{\tt T}\right)$.
    Here, $\gamma_{i j}$ represents the transport plan indicating how much mass is transported from sample $\mathbf{H}_{{\tt S}} ^i$ in the source domain $\mathcal{D} _{\tt S}$ to sample $\mathbf{H}_{{\tt T}} ^j$ in the target domain $\mathcal{D} _{\tt T}$,
$\left\Vert \cdot \right\Vert _F$ denotes the Frobenius norm, which is defined as
    $\|\mathbf{A}\|_F= \sqrt{\sum_{i, j}\left|a_{i j}\right|^2}=\sqrt{\operatorname{tr}\left(\mathbf{A}^* \mathbf{A}\right)}$,
where $\mathbf{A}^*$ is the conjugate transpose of matrix $\mathbf{A}$, and
 $\operatorname{tr}$ denotes the trace.
In our experiments, we have 3077 samples in the QSCM dataset and 1000 samples in the MBCM dataset.
    Due to the large size of our datasets, we need the Sinkhorn approximation \cite{Chen2021} to approximate our Wasserstein-1 distance.
The Wasserstein-1 distance with Frobenius norm between our two datasets is approximated to be $0.4597$, which seems to be a relatively large value, considering that our channel magnitudes are in the range of $[0, 3\times 10^{-5}]$ and $[0, 12\times 10^{-5}]$.
    This demonstrates that the two domains are significantly different.

\subsection{Target Model}
    In the model architecture applied to the target domain, the early layers are kept frozen and remain untrainable.
The parameters of these frozen layers, inherited from the source domain, are denoted as $\theta_{\tt S} ^{\tt frozen}$, 
    and these weights are not modified during the fine-tuning process.
    
{In the case of CNNs, the last three layers are trainable, while the earlier layers remain frozen.
  Specifically, the frozen layers are denoted as $g_i^{\tt conv, f}, g_i^{\tt act, f} (i\in \{ 0,...,4\})$, 
    meaning that $g_i^{\tt conv, f}, g_i^{\tt act, f}$ in the target domain is identical to $g_i^{\tt conv}, g_i^{\tt act}$ from the QSCM model,  
      while $g_i^{\tt conv, t}, g_i^{\tt act, t} (i \in \{5,6,7\})$ denote the trainable layers.}

\subsection{Training in Target Domain}

    The parameters of the non-frozen layers are denoted as $\theta_{\tt T} ^{\tt trainable}$, which are the only weights updated during fine-tuning. 
The fine-tuned parameters can be obtained by solving the following optimization problem (see Eq. \eqref{eq:theta_trainable1}).
\begin{align}
    \theta_{\tt T} ^{\tt trainable, \ast} = \arg \min_{\theta_{\tt T}} \mathcal{L}_{\tt T}\left( \theta_{\tt T} \right).
    \label{eq:theta_trainable1}
\end{align}
Combining the fixed and trainable parameters, the model for domain adaptation can be expressed as shown in Eq. \eqref{eq:theta_trainable2}.
\begin{align}
    \theta_{\tt T} ^{\tt {trainable, \ast }}
     =  \arg \min_{\theta_{\tt T}^ {\tt {trainable }}} 
    F \left( \theta_{\tt S} ^{\tt {frozen }}, \theta_{\tt T}^{\tt {trainable }}\right),
    \label{eq:theta_trainable2}
\end{align}
with 
    $F (x_1,x_2) = 
    \mathbb{E}_{ \left( \mathbf{ \tilde{H}}_{\tt T}, \mathbf{H}_{\tt T} \right) \sim P_{\tt T}} \left[ \ell \left( f\left( \mathbf{\tilde{H}} _{\tt T} ; x_1, x_2\right), \mathbf{{H}}_{\tt T} \right) \right]$,
    where $f\left( \mathbf{\tilde{H}} _{\tt T}; \theta_{\tt S} ^{\tt {frozen }}, \theta_{\tt T}^{\tt {trainable }}\right)$ is the output of the model on the target domain data $\mathbf{\tilde{H}}_{\tt T}$, 
      using the frozen source domain parameters $\theta_{\tt S} ^{\tt frozen}$ 
      and the trainable parameters $\theta_{\tt T} ^{\tt trainable}$.

    The results obtained from the CNN and GAN models can be expressed as $f\left( \mathbf{\tilde{H}} _{\tt T}; \theta_{\tt S} ^{\tt {frozen }}, \theta_{\tt T}^{\tt {trainable, \ast }} \right)$.
On de-normalizing these results with $\min$, $\max$ of $\mathbf{\hat{H}}^{\tt ls}$ and $\mathbf{\hat{H}}^{\tt li}$, we get estimated channels $\mathbf{\hat{H}}^{\tt ls, cnn}$, $\mathbf{\hat{H}}^{\tt li, cnn}$ and $\mathbf{\hat{H}}^{\tt ls, gan}$, $\mathbf{\hat{H}}^{\tt li, gan}$ in the target domain, respectively.

Similarly, in GANs, the early layers of both the generator and the discriminator are frozen.
    In the generator network, layers $f ^{\tt G} _{1, \tt enc}, ..., f ^{\tt G} _{5, \tt enc}$ are frozen, 
  while in the discriminator network, layers $f ^{\tt D} _{1}, f ^{\tt D} _{2}, f ^{\tt D} _{3}$ are frozen.
We refer to these models as MBCM models.

\section{Numerical Results}
{In this study, we use two different domains to generate our channels.
    The first domain is the QSCM 
where the channels are generated according to Eq. \eqref{eq:H_km}.
    The second one is the MBCM, which uses the map of the \'{E}TS neighborhood.
    The channels in the MBCM are generated using MATLAB CDL Channel Model with RayTracing approach. 
The first domain serves as a simulation framework, so we call it the source domain, while the second domain closely resembles a practical system, so we call it the target domain.
}

{For the QSCM, we use DeepMIMO with the Outdoor 1 scenario with carrier frequency of $3.4$ GHz and Eq. \eqref{eq:H_km} to generate the channels. 
    The configuration includes BS 16 and UEs from rows 3500 to 3516, yielding a total of 3077 samples.}

{In order to generate the MBCM dataset, we first import the .osm file of \'{E}TS neighborhood map from OpenStreetMap (OSM). 
Next, we set the locations of BS and UEs on the map. 
    Using the RayTracing object in MATLAB, we generate the CDL channels between the BS and UEs.}
 The system adopts OFDM in the 5G-NR setting, based on Table 4.3.2-1 in technical specifications TS-38.211 \cite{TS.38.211}, 
    with numerology $\mu = 1$, subcarrier spacing (SCS) of $30 $kHz, $M=14$ symbols per slot, $10$ slots per sub-frame, and $2$ sub-frames per frame. 
We consider a system with $K = 612$ subcarriers and define a Resource Grid (RG) for one slot consisting of a total of $612$ subcarriers and $14$ OFDM symbols.

\begin{figure}[!htbp]
    \centering
    \includegraphics[width = .8\linewidth]{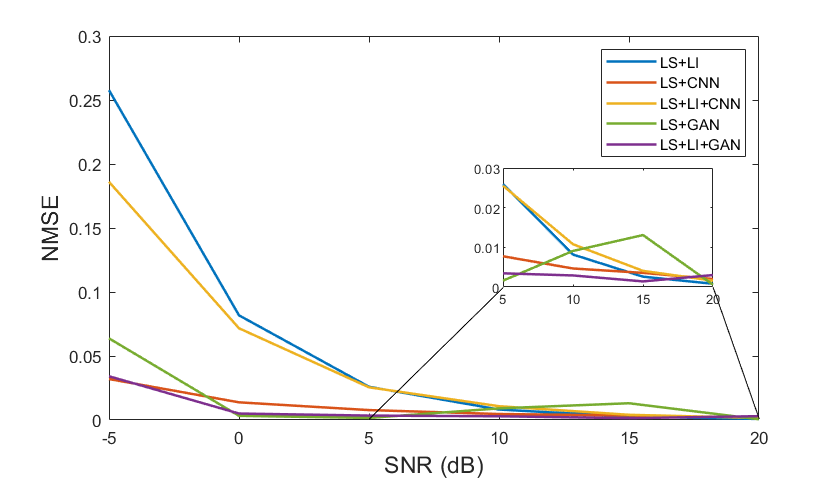}
    \caption{Performance of trained models on source domain using QSCM-based generated dataset.}
    \label{fig:4_NMSE_eval}
\end{figure}

On training our models on the QSCM dataset (the source domain), we first evaluate their performance on this domain. 
    The results for the CNN-based and GAN-based models are compared to the traditional LS-LI method (see Fig. \ref{fig:4_NMSE_eval}). 
The results indicate that our CNN-based and GAN-based models offer significant improvements on low SNR levels, while at high SNR, the LS-LI method performs sufficiently well and does not require further refinement from the CNN model.
    {The LS-LI-CNN approach, however, is less effective than the others, likely due to the errors introduced by the LS-LI method, which the CNN is unable to correct.}

\begin{figure}[htbp]
    \centering
    \includegraphics[width = 0.8\linewidth]{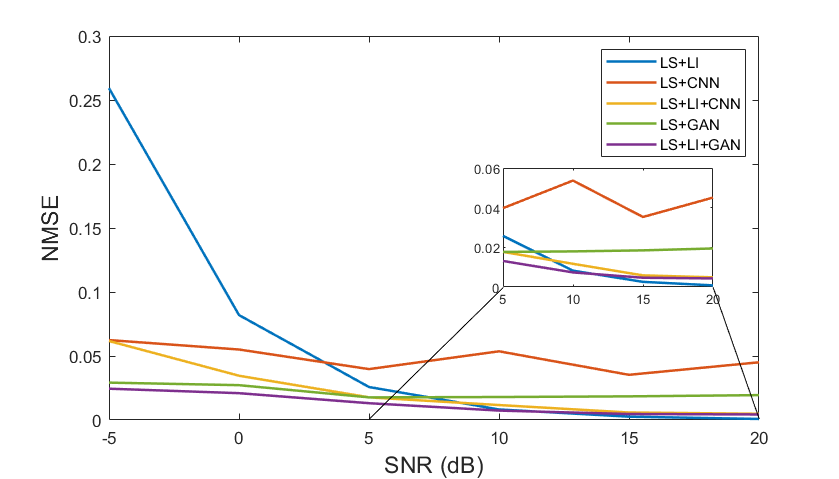}
    
    \caption{Performance of transfer learning on the target domain using OSM-based generated dataset.}
    \label{fig:6_transfer_NMSEv5}
\end{figure}

In the target domain, represented by the MBCM dataset, we generate 1000 samples. 
    From these, we use only 300 samples for fine-tuning and the remaining 700 for validation. 
This setup closely reflects a practical scenario where limited ground-truth channel data for model training are available. 
    A comparison of the performance of our fine-tuned machine-learning models and the LS-LI estimation method is shown in Fig. \ref{fig:6_transfer_NMSEv5}.
The results of our fine-tuned models demonstrate significant improvements in low-SNR conditions, while at high SNR, the LS-LI approach performs well without requiring further enhancement. 
    Although the LS-CNN and LS-GAN models perform well in the source domain, they turn out
        to be less effective in the target domain. 
This indicates that, compared to the LS-CNN and LS-GAN models, the LS-LI-CNN and LS-LI-GAN models more efficiently adapt to the target domain during fine-tuning.

\begin{figure}[htbp]
    \centering
    \includegraphics[width = 0.8\linewidth]{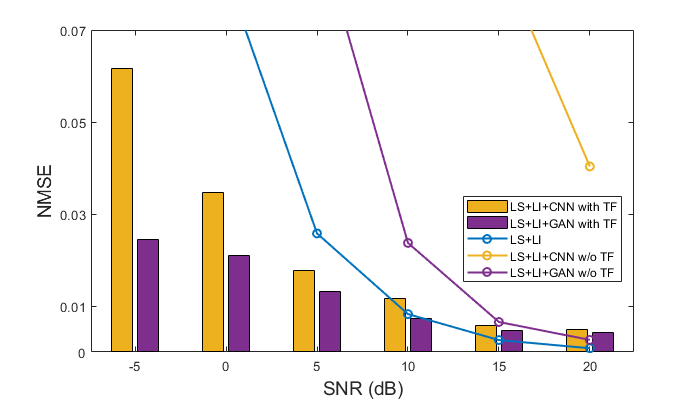}
    \caption{Comparison between approaches with and without transfer learning in the target domain.}
    \label{fig:6_compare5_2}
\end{figure}

Fig. \ref{fig:6_compare5_2} presents a performance comparison between the LS-LI-CNN and LS-LI-GAN models, both with and without fine-tuning in the target domain. 
    Notably, without fine-tuning, the ML-based models underperform compared to the LS-LI approach, indicating that instead of mitigating errors, they exacerbate the inaccuracies introduced by the LS-LI method. 
However, once fine-tuned, these models exhibit substantial performance improvements, demonstrating their ability to adapt and enhance accuracy in the target domain.

\section{Conclusion}

This study applies CNN- and GAN-based models for channel estimation in OFDM systems, showing their effectiveness in low SNR conditions, while LS-LI remains competitive at higher SNRs. 
    A key contribution is leveraging transfer learning to adapt models from simulation-based datasets to real-world scenarios, significantly improving performance by addressing domain shift and enhancing generalization in practical channel estimation.

Future research may extend the proposed framework to address time-varying channels, where the channel can change rapidly within an OFDM slot. 
    Furthermore, unsupervised domain adaptation techniques will be explored to handle practical scenarios in which ground truth labels in the target domain are unavailable, thereby enhancing the model's real-world applicability.


\end{document}